# ELECTRICAL AND OPTICAL PROPERTIES
## OF SEMICONDUCTORS

# Specific Features of the Luminescence and Conductivity of Zinc Selenide on Exposure to X-Ray and Optical Excitation

## V. Ya. Degoda^ and A. O. Sofienko

*Kyiv National University, Physical Department, Kyiv, 03680 Ukraine*
*^e-mail: degoda@univ.kiev.ua*
Submitted September 8, 2009; accepted for publication September 15, 2009

**Abstract**—The set of experimental data on the X-ray-excited luminescence and X-ray induced conductivity of ZnSe are compared to the data on the photoluminescence and photoconductivity. It is experimentally established that the current—voltage characteristics and the kinetics of phosphorescence and current relaxation depend on the type of excitation. It is found that the external electric field influences the intensity and shape of bands in the luminescence spectra. It is shown that the character of excitation defines the kinetics of recombination, charge carrier trapping, and conductivity in wide-gap semiconductors.

**DOI:** 10.1134/S106378261005@@@@

## 1. INTRODUCTION

Zinc selenide belongs to the most promising wide-gap II—VI materials. This material finds wide use in the development of devices for short-wavelength semiconductor electronics and displaying systems [1—4] and offers promise for electronics of light-emitting-diodes [5, 6]. Other fields of application of single-crystal ZnSe are associated with the use of the material for detectors of ionizing radiation. This became possible after the development of technologies of growth of rather perfect crystals with low concentrations of uncontrollable impurities and a high resistivity (at a level of $10^{10}$—$10^{11}$ $\Omega$ cm) [7, 8]. The large average atomic number ($Z = 32$) and the wide band gap (2.7 eV at 300 K [9]) make ZnSe promising for the production of X-ray detectors that have no need of cooling. It was experimentally established that, on detection of ionizing radiation (e.g., X-ray radiation [10]), the kinetics of conductivity and luminescence in ZnSe substantially differed from the corresponding kinetics under optical excitation and no explanations of this observation in the context of the classical theory of photoconductivity and photoluminescence (PL) in semiconductors were offered [11, 12]. In addition, it was noted that, even in undoped crystals, an external electric field had an effect on the PL and X-ray-excited luminescence (XRL). Therefore, the purpose of this study was to conduct a series of experimental investigations of the luminescence, conductivity, phosphorescence (Ph), and current relaxation in ZnSe on X-ray and optical excitation and to clarify the basic effects responsible for the dependence of the kinetics of luminescence and conductivity in ZnSe on the type of excitation.

## 2. EXPERIMENTAL

The luminescence and conductivity of the ZnSe single crystals were studied under conditions of band-to-band excitation with optical and X-ray photons. In order to produce crystals with the lowest concentration of impurities and the highest resistivity, we grew the nominally undoped ZnSe samples after purification of the working mixture. For conductivity measurements, we deposited electrical indium contacts onto the single crystals by the resistive method and then soldered leads to the contacts. The spacing between the electrodes was 5 mm. A voltage no higher than 1000 V was applied to the electrodes, one of which was grounded via a nanoammeter. The input impedance of the nanoammeter was much (several orders of magnitude) lower than the resistance of the ZnSe sample. The experimental studies were carried out in the temperature range from 85 to 450 K. The X-ray excitation of the samples was accomplished with the integrated emission of a BKhV7 X-ray tube (Re, 20 kV, 5—25 mA), through the beryllium window of the cryostat. The distance from the anode of the X-ray tube to the sample was 120 mm. For optical excitation, we used seven ultraviolet light-emitting diodes (LEDs) UF-301 (emitting at the wavelength 395 nm with the current 5—30 mA) that simultaneously illuminated the sample through the quartz window of the cryostat. The radiation intensities of the X-ray tube and LEDs are proportional to the supply current, whereas the shape of their spectra remains unchanged in this case. The luminescence signal was transmitted through an MDR-2 monochromator and then detected with the use of the FÉU-106 and FÉU-83 photomultipliers (operating on cooling). The spectra were corrected for the spectral sensitivity of the recording system. The emission spectra and conductivity spectra were recorded simultaneously. At the same time, we also recorded the phosphorescence and current relaxation after cessation of excitation.

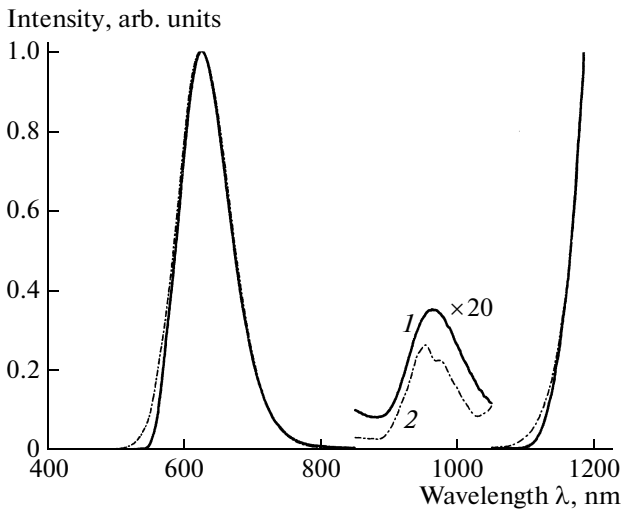

**Fig. 1.** Normalized luminescence spectra of ZnSe on (*1*) optical and (*2*) X-ray excitation. $T = 85$ K.

The luminescence excitation spectra and the photoconductivity spectra were studied with the use of a KGM-150 (24 VGF, 6 A) filament lamp with a continuous spectrum and an MDR-2 monochromator with quartz condensers. The monochromatic radiation was directed to the area of the sample between the contacts. The luminescence signal was recorded through the quartz lens window in the direction orthogonal to the monochromatic excitation radiation. The spectra were corrected for the intensity of the monochromatic radiation incident on the sample.

## 3. RESULTS AND DISCUSSION

### 3.1. Emission Spectra of ZnSe

In the XRL and PL spectra of the ZnSe single crystals, we observe broad recombination emission bands with peaks at the wavelengths 626 and 963 nm. At low temperatures, the edge emission and the short-wavelength emission associated with donor–acceptor pairs, with the photon energies close to the band gap, are practically unobservable. This is typical of polycrystalline samples or doped single crystals [9]. The lack of the contribution of donor–acceptor pairs to the emission spectra of the samples is indicative of a low concentration of uncontrollable impurities, and the very low intensity of edge emission at the temperature $T = 85$ K can be attributed to pronounced temperature quenching of this luminescence. The emission band prevailing in the spectrum is the self-acti-vated luminescence band at the wavelength $\lambda = 626$ nm (1.98 eV). The authors of [9] attribute this band to the complex center $V_{Zn} + D$ that consists of a Zn vacancy and a shallow impurity. It is known that the luminescence spectra of ZnSe involve also an infrared emission band with a peak at $\lambda = 963$ nm (1.29 eV) [9, 13]; this band is related to vacancies of selenium $V_{Se}$. Figure 1 shows the normalized luminescence spectra of single-crystal ZnSe. An important feature of the XRL and PL spectra is their different temperature behaviors. If the PL excitation intensity is chosen so that the PL and XRL intensities at room temperature are at the same level, the PL and XRL intensities at low temperatures ($T = 85$ K) differ by a factor close to 10: the PL is almost ten times more intense. This suggests that the temperature quenching of luminescence is controlled not only by the intraenter quantum yield of emission centers, but also by the probability of localization of charge carriers opposite in sign at a point defect that serves as an emission center. If the integrated XRL intensity of the ZnSe single crystals is compared to the almost an order of magnitude higher integrated XRL intensity of the classical ZnSe–Cu phosphor, it becomes obvious that, at practically identical rates of generation of free charge carriers, there is also noticeable nonradiative recombination that occurs in ZnSe along with radiative recombination.

On application of a potential difference to the electrodes of the crystal, the shape of the bands in the emission spectra of ZnSe changes and, at the same time, the integrated luminescence intensity decreases. Such a trend can be clearly seen in the ZnSe samples on optical and X-ray excitation exactly at room temperature, whereas on cooling of the sample to 85 K, the effect of the external field becomes much less pronounced. In the case of XRL, an increase in the field yields also a broadening of the prevailing recombination band at 626 nm. In the case of PL, the broadening of the emission band is not so noticeable and only a change in the intensity of the band is clearly observed.

The ratio of the luminescence intensity of the prevailing 626-nm band in the spectrum with no external field to the intensity in a nonzero field characterizes the changes in the spectra. Figure 2 shows the spectra of such a ratio in the cases of PL and XRL.

### 3.2. Photoconductivity and X-Ray-Excited Conductivity of ZnSe

The conductivity of ZnSe has rather long been studied for both undoped and doped samples [9, 12]. Zinc selenide can be easily doped with shallow donors (Al, Ga, In) that provide the $n$-type conductivity even at liquid-helium temperatures. The optically perfect undoped ZnSe crystals exhibit a low dark conductivity (the current is below 50 pA at 300 K) that is due to thermally activated delocalization of charge carriers from a deep level (0.31 eV). In this case, a change in the arrangement of the contacts (their area or the spacing between them) influences only the amplitude of the current, whereas the slope of the dark conductivity plotted as $\ln(I)$ versus $10^3/T$ or the activation energy defining the slope remains unchanged.

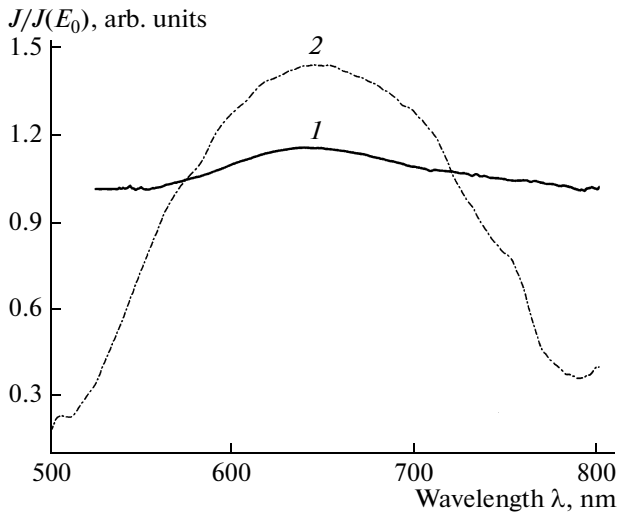

**Fig. 2.** The ratio of ( *1* ) the PL and ( *2* ) XRL intensities measured for the 626-nm (1.98 eV) emission band in zero field, $J$, to the corresponding emission intensities measured in the electric field, $J(E_0)$. $E_0 = 600 \text{ V cm}^{-1}$.

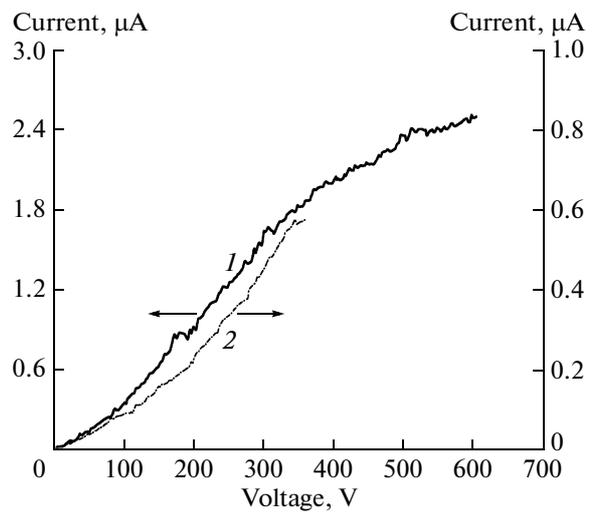

**Fig. 3.** Current–voltage characteristics of the X-ray-excited conductivity of ZnSe at ( *1* ) 295 and ( *2* ) 85 K.

The conductivities observed on optical and X-ray excitations are noticeably (by several times) different at the PL and XRL intensities close to each other. The current–voltage ($I$–$V$) characteristics recorded on X-ray excitation of the sample at different temperatures are shown in Fig. 3. For the single crystals at the temperature $T = 295$ K, the mobility of free charge carriers is much lower than the mobility at $T = 85$ K [11], which is due to prevailing scattering at acoustic phonons of the crystal lattice. With increasing electric field in the sample up to 1000 V cm$^{-1}$, we observe the steadily increasing current. In the case of X-ray-excited conductivity, the $I$–$V$ characteristic is nonlinear. For the photoconductivity, the $I$–$V$ characteristics is superlinear (Fig. 4). As the temperature of the sample is changed from 295 to 85 K, the X-ray-excited conductivity decreases, whereas the photoconductivity increases. At the same time, the PL and XRL intensities always increase with decreasing temperature. Consequently, the value of the X-ray-excited conductivity is controlled not only by the mobility of charge carriers generated on absorption of X-ray photons, but also by the temperature dependence of the recombination rate of charge carriers.

## 4. JUMPS OF THE LUMINESCENCE INTENSITY AND CONDUCTIVITY ON TERMINATION OF EXCITATION

On excitation of a phosphor crystals, the PL intensity is defined by the concentration of recharged emission centers and the concentration of free charge carriers that recombine at these centers [11]. Since charge carriers are in the localized state most of the time, the processes of charge carrier generation, recombination, and trapping at and delocalization from traps become balanced within some time interval after the beginning of illumination. Balance equations provide a means for determining the steady-state concentration of free

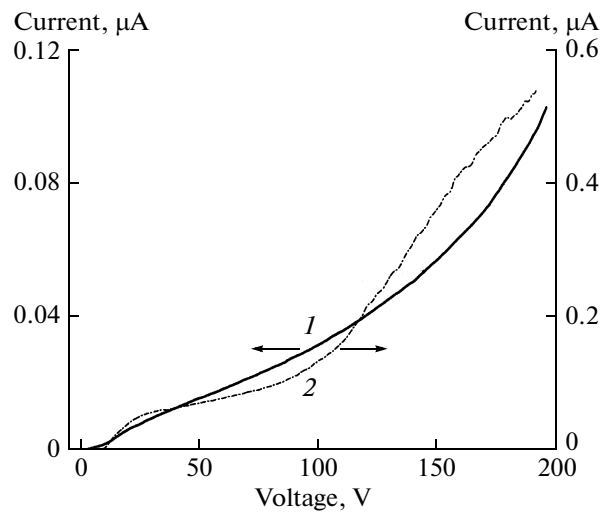

**Fig. 4.** Current–voltage characteristics of the photoconductivity of ZnSe at ( *1* ) 295 and ( *2* ) 85 K.

Intensity, arb. units

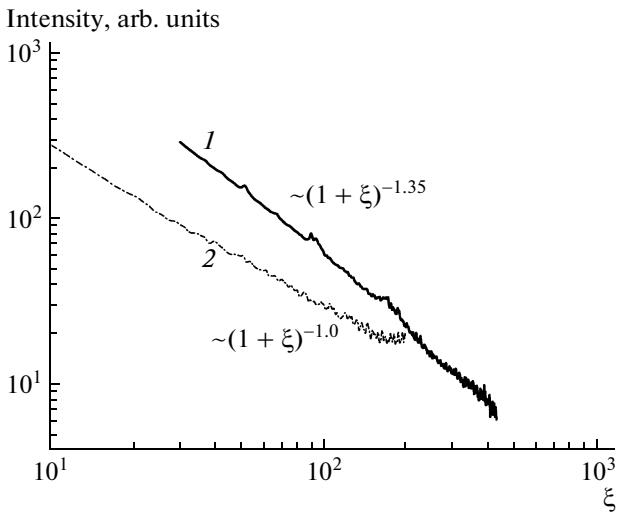

**Fig. 5.** The phosphorescence decay curves recorded at the wavelength 626 nm after termination of (*1*) optical and (*2*) X-ray excitation. $T = 85$ K.

charge carriers in the conduction band, $N$, for the simple model of a phosphor crystal [11]:

$$N = \frac{k_0 I_0 + \sum_i w_i n_i}{\sum_k \sigma_{0k} n_k^+ + \sum_i \sigma_i \nu_i}.$$

Here, $k_0 I_0$ is the intensity of generation of free charge carriers in the crystal; $w_i n_i$ is the intensity of delocalization of charge carriers from traps (the product of the probability of delocalization per unit time by the concentration of occupied traps); $\sigma_{0k}$ is the recombination cross section of the $k$-type emission centers, the concentration of which is $n_k^+$; and $\sigma_i$ is the cross section of localization at the traps, the concentration of which is $\nu_i$. At the time point of termination of excitation, the concentrations of recharged emission centers and traps remain at their steady-state levels, whereas the concentration of free carriers directly involved in recombination drastically decreases. This results in an abrupt (by a factor of $10-1000$) jump of the emission intensity. Based on the balance equations, the jumps of the conductivity and luminescence intensity are bound to be identical, since these jumps are defined by the change in the concentration of free charge carriers. Thus, the measurement of the jumps of the conductivity and luminescence intensity allows one to compare the contribution of generated charge carriers to the concentration of free carriers with the contribution of charge carriers delocalized from traps.

In the case of optical excitation, we observe a clearly pronounced correlation between the jumps of the luminescence intensity and conductivity: the jumps are of the same order of magnitude and correspond to a factor of $\sim 2 \times 10^3$. This result does not contradict the above-described conclusions of the classic kinetic theory of PL. However, upon X-ray excitation, the jump of the luminescence intensity is about half the jump on optical excitation and corresponds to the factor $10^3$, whereas, for the X-ray-excited conductivity, the jump is several orders of magnitude smaller than the jump on optical excitation and corresponds to a factor of 1.5. This result cannot be explained only on the basis of statements of the classical kinetic theory, since the above observations suggest that the major contribution to the X-ray-excited conductivity is made precisely by such charge carriers that are delocalized from deep traps outside the excited local regions. Most charge carriers generated by X-ray excitation are bound to recombine rather fast in the region of generation, so that they have no time to contribute to the total X-ray-induced conductivity of the crystal. At the same time, we believe that the majority of generated charge carriers recombine without localization at deep traps, as supported by the low-intensity thermostimulated luminescence (TSL); therefore, after termination of excitation, the jump of the luminescence intensity is bound to be rather substantial and exceed the jump observed on termination of optical excitation of the sample. However, such a relation is not observed experimentally.

The above-described results show a substantial difference in the kinetics of trapping of charge carriers generated by different types of excitation. This may be due to the high concentration of electronic excitations in the bulk of the crystal on absorption of an X-ray photon.

## 5. PHOSPHORESCENCE AND CONDUCTION CURRENT RELAXATION IN ZnSe

The phosphorescence decay kinetics and the current relaxation kinetics after termination of optical and X-ray excitation of single-crystal ZnSe at the temperature 85 K are described by the hyperbolic time dependence. The typical phosphorescence decay curves recorded at the wavelength 626 nm are shown in Fig. 5. Such temperature of measurements was chosen, since there is no accumulation of the light sum at room temperature. In addition, it should be noted that the TSL peak intensities are several orders of magnitude lower than the steady-state luminescence intensity, as for the thermostimulated conductivity (TSC) peaks that are much lower that the steady-state conductivity. In the TSL and TSC spectra, peaks are observed in the range $100-250$ K typical of ZnSe. Since the phosphorescence intensity is proportional to $(1 + wt)^{-\alpha}$ (where $t$ is time and $w$ is the probability of thermal delocalization of charge carriers from traps), all of the phosphorescence decay and current relaxation curves are plotted at the dimensionless time scale $\xi = wt$. This allows us, first, to linearize the curves on the log–log scale and determine $w$ and, second, to simplify the approximation of the curves, since only one basic unknown parameter—specifically, the exponent of the hyperbola—remains in this case. In the case of X-ray excitation, the phosphorescence decay curves can be linearized in the log–log scale with the same values of $w$, as in the case of optical excitation; consequently, trapping of charge carriers occurs at the same centers. In contrast to what happens on termination of optical excitation, on termination of X-ray excitation, we observe a rather slow decay of the emission intensity and the experimental phosphorescence decay curves cannot be interpreted in the context of classical approaches to the PL kinetics. To explain such a slow phosphorescence decay, the authors of [14] consider the contributions of individual local excitation regions formed on absorption of X-ray photons to the total phosphorescence intensity of the sample. These local regions characterized by high concentrations of occupied traps and ionized emission centers are created during the entire time of excitation. To perform a detailed calculation of the phosphorescence decay curves, it is necessary to know the kinetics of recombination and retrapping of charge carriers in the place of absorption of an X-ray photon. This presents a rather difficult problem. It is obvious that such a macroscopically and microscopically nonuniform spatial distribution of the occupancy of traps predetermines modifications also in the current relaxation kinetics on X-ray excitation. In particular, such nonuniform distribution is responsible for the slower decay of the conduction current with time. Figure 6 shows the current relaxation curves recorded after termination of optical and X-ray excitation. It should be noted that, in contrast to the case of optical excitation, in the case of X-ray excitation of ZnSe, the current relaxation is faster than the phosphorescence decay ($\alpha = 1.1$ and 1.0). This experimental fact is essentially impossible to explain in the context of the classical kinetic theory of PL [11].

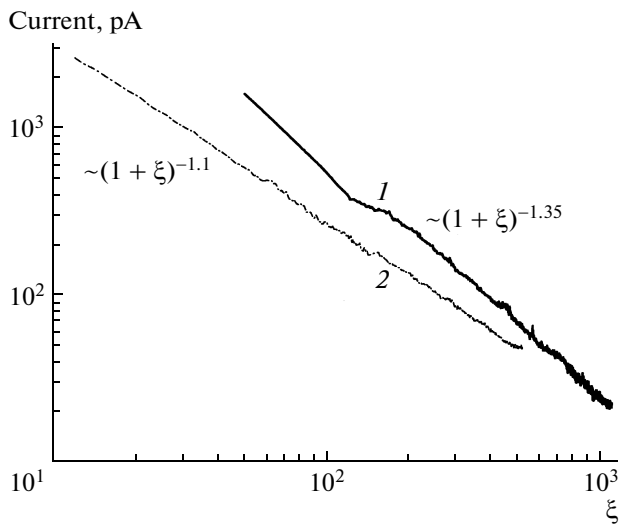

**Fig. 6.** Conduction current relaxation curves recorded after termination of (*1*) optical and (*2*) X-ray excitation. $T = 85$ K.

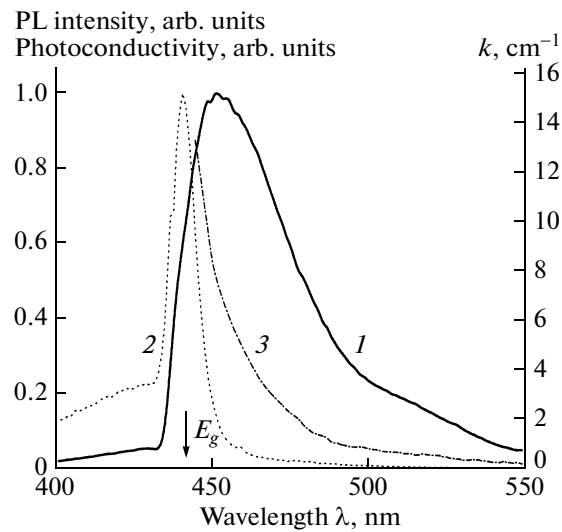

**Fig. 7.** The (*1*) PL and (*2*) photoconductivity excitation spectra and the (*3*) spectrum of the absorption coefficient $k$ of ZnSe at 85 K.

Since the phosphorescence decay and current relaxation curves were recorded simultaneously, an additional electric field was applied to the sample. To study the effect of this electric field on the phosphorescence and current relaxation kinetics, we recorded the decay curves after excitation during equal time intervals at different voltages applied to the electrodes of the sample. For the current relaxation, we observe a well-pronounced correlation between the exponent of the hyperbola that describes the current decay and the potential difference applied to the electrodes. The relative variation in the exponent of the hyperbola corresponds to +0.36% V⁻¹. The effect of the electric field on the phosphorescence decay is less pronounced: we observe the tendency for a slight (−0.11% V⁻¹) decrease in the exponent of the hyperbola. Such opposite changes in the phosphorescence decay and current relaxation with electric field invite further investigations and analysis.

## 6. ABSORPTION, LUMINESCENCE EXCITATION, AND PHOTOCONDUCTIVITY SPECTRA OF ZnSe

To clarify the issue of why the quantum yield of recombination luminescence in ZnSe is relatively low in comparison to the quantum yield in ZnS–Cu, we recorded the PL excitation, photoconductivity, and transmittance spectra of ZnZe in the wavelength range from 350 to 550 nm (Fig. 7). The photoconductivity spectra were recorded simultaneously with the luminescence excitation spectra. At room temperature, the spectral positions of the peaks in the PL and photoconductivity spectra coincide with each other and are close to the fundamental absorption edge (461 nm); the widths at half maximum of the bands are practically the same (~12 nm). On cooling the sample to 85 K, the peak of the photoconductivity spectrum shifts to shorter wavelengths together with the band gap (the spectral width of the photoconductivity curve remains practically unchanged) and additional broad bands appear in the PL excitation spectrum (Fig. 7). The luminescence excitation spectrum can be decomposed into three bands, of which one (short-wavelength) band coincides with the photoconductivity spectral band, the two other bands apparently being the bands of excitation of intracenter luminescence. It may be suggested that these two bands correspond to impurity excitation or excitons that cannot be clearly seen at room temperature.

The transmittance spectra of ZnSe show that the electric field below 500 V cm$^{-1}$ has no effect on the transmittance and the spectral position of the basic peaks in the luminescence excitation and photoconductivity spectra at room temperature correspond to the energy higher than the band gap $E_g$. With decreasing temperature, the photoconductivity peak shifts together with $E_g$ by the same energy.

At room and liquid-nitrogen temperatures, we observe a sharp decrease in the PL intensity and photoconductivity at the energy of excitation photons $E_g + 0.050$ eV. It may be thought that such behavior is caused by two factors, one of which is defined by uncontrollable nonradiative recombination centers, at which free charge carriers can be localized only if they overcome a potential barrier, and the other of which may be associated with surface nonradiative recombination centers. At higher absorption coefficients $k$, the absorption occurs closer to the surface, where the concentration of nonradiative recombination centers is very high. The first of the above-mentioned factors is supported by the XRL intensity. In fact, free charge carriers generated by X-ray excitation are hot and can easily recombine at nonradiative recombination centers with barriers, resulting in a low XRL quantum yield compared to the quantum yield of the classical ZnS−Cu phosphor.

## 4. CONCLUSIONS

From the results of comprehensive experimental studies of the luminescence and conductivity of zinc selenide, a number of conclusions can be drawn. At equal numbers of free electron−hole pairs generated in the ZnSe single crystals on absorption of optical and X-ray photons, different luminescence intensities and conductivities are observed for these two types of excitation. It is found that the photoconductivity and X-ray-excited conductivity are substantially different in $I-V$ characteristics and their temperature behavior.

With increasing external electric field to 1000 V cm$^{-1}$, the luminescence intensity decreases, irrespective of the type of excitation radiation. On X-ray excitation of ZnSe, a noticeable broadening of the emission band at the wavelength 626 nm is observed in the electric field. The effect suggests that, on X-ray excitation, the electric field noticeably influences the recombination processes because of the well-pronounced spatial nonuniformity of the distribution of free charge carriers when generated.

The exponents of hyperbolas that describe the phosphorescence decay and conduction current relaxation after termination of optical or X-ray excitation are substantially different for these two types of excitation. This difference cannot be accounted for in the context of the classical theory of luminescence in phosphor crystals, suggesting that the microscopically nonuniform excitation of the crystal by X-ray photons has a profound effect on the kinetics of trapping of charge carriers.

A noticeable decrease in the PL intensity and photoconductivity is observed at the excitation photon energy $E_g + 0.05$ eV. This may be due to a high concentration of nonradiative recombination centers.

Thus, comprehensive experimental studies of the luminescence and conductivity of ZnSe on optical and X-ray excitation show that not all results can be interpreted in the context of the classical kinetic theory of PL and photoconductivity even at a qualitative level. To accomplish quantitative comparative analysis of the data, it is necessary to develop a kinetic theory of X-ray-excited luminescence and conductivity with consideration for the substantially nonuniform spatial generation of electronic excitations in solids.

*Translated by É. Smorgonskaya*

SPELL: OK